\documentclass{article}

\usepackage{PRIMEarxiv}

\usepackage[utf8]{inputenc} % allow utf-8 input
\usepackage[T1]{fontenc}    % use 8-bit T1 fonts
\usepackage{hyperref}       % hyperlinks
\usepackage{url}            % simple URL typesetting
\usepackage{booktabs}       % professional-quality tables
\usepackage{amsfonts}       % blackboard math symbols
\usepackage{nicefrac}       % compact symbols for 1/2, etc.
\usepackage{microtype}      % microtypography
\usepackage{multirow}
\usepackage{booktabs}
\usepackage[ruled,vlined,linesnumbered]{algorithm2e}
\usepackage{amsmath}
\usepackage{fancyhdr}       % header
\usepackage{graphicx}       % graphics
\graphicspath{{media/}}     % organize your images and other figures under media/ folder

\title{Forward asymmetric numeral systems coding for natural language text compression
}

\author{
  M. Kharin, I. Zavadskyi \\
  Faculty of Computer Science and Cybernetics \\
  Taras Shevchenko National University of Kyiv \\
  Kyiv, Ukraine\\
  \texttt{nikitaolimp2003@gmail.com, ihorzavadskyi@knu.ua} \\
}

\begin{document}
\maketitle

\begin{abstract}
Compression based on asymmetric numeral systems (ANS) combines high encoding and decoding speeds with a compression ratio close to Shannon entropy, while forward modeling of the information source makes it possible to obtain an estimated compressed message size that is less than the entropy. This paper proposes combining these modeling and coding methods. In addition to ensuring high data processing speeds and compression ratios, this approach enables one to implement the adaptive ANS, which has long remained an important scientific and practical problem.
\end{abstract}

% keywords can be removed

\section{Introduction}
Data compression remains one of the key problems in information processing. For a static information source without memory, the theoretical bound for its compression level is determined by its entropy, which can be computed using the Shannon formula. Arithmetic Coding (AC) \cite{rissanen1976} guaranties that the compression level is approximately equal to the bound; however, it is characterized with high computational complexity. \\
Relatively recently, a new method based on Asymmetric Numeric Systems (ANS) has been presented \cite{duda2009, duda2014}, which has drastically faster performance and not significantly inferior compression ratio than AC. It uses \textit{a state}, a natural number, which is changed almost inversely proportionally to the probability of a symbol on each en(de)coding step. One of the possible implementations of ANS is tabled ANS (tANS), which is based on a table of a permutation of the input indices and guaranties the compressed size to be $H + O(m)$ bites, unlike ranged ANS (rANS), which produces $H + O(n)$ bites, where $m$ and $n$ are the sizes of the dictionary and input text respectively, and $H$ is the entropy. \\
One of its features is backward decoding, which makes common adaptive coding based on accumulative frequencies impossible, since the decoder will not know the character's frequencies, which were used by the encoder. However, this problem can be resolved with Forward Adaptive Modeling (FAM), which computes the original frequencies at the beginning and reduces them during text processing \cite{zavadskyi2024}. The theoretical lower bound of the output code, conveyed with FAM, is less than the entropy \cite{zavadskyi2026}. \\
We propose an adaptive realization of ANS based on Forward Modeling, which surpasses the original ANS in terms of compression ratio, in particular, for natural language text compression.

\section{Algorithm and implementation details}
\label{sec:headings}

We will describe the general idea of the method. Using FAM for encoding in the forward direction, we decrease character's frequencies from the original ones to zero. The decoder operates in the backward direction, using a common adaptive approach, i.e. increasing the frequencies starting from zero. Thus, one does not have to transmit them to the decoder. The problem of adding symbols that have not appeared in the input is resolved by sorting the dictionary in the order of the last appearance and replacing them with an auxiliary symbol $lt$. Since the decoder works backward, with reading a $lt$ it takes the corresponding symbol from the dictionary and assigns $1$ to its frequency. \\
The important detail which determines the differences of tANS implementations is a way to permute the characters in the table. We claim that for large alphabets the symbols may appear in the table in the same order as in the text (or vise versa). Experiments confirm the efficiency of this approach. \\
Let $T$ denote the input text, $n = |T|$ its size, $W_0$ the dictionary, and $\text{Code}$ the output bit stream. Now we can formulate the algorithm. \\

\begin{algorithm}[H]

\SetAlgoLined

\KwIn{$T$}
\KwOut{$\text{Code}, W_0$}

Construct $W_0$ from $T$ (sorted by last occurrences)\;

Construct modified sequence $T'$ by inserting $\text{lt}$ after the last occurrence of each token in $W_0$, then reverse $T'$\;

\ForEach{$t \in W_0 \cup \{\text{lt}\}$}{
    $I_t \leftarrow \{\, i \mid T'[i] = t \,\}$\;
    \eIf{$t \in W_0$}{
        $f[t] \leftarrow |I_t| - 1$\;
    }{
        $f[t] \leftarrow |W_0|$\;
    }
}

$x \leftarrow n + |W_0|$\;
$l \leftarrow n + |W_0|$\;

\For{$i \leftarrow 0$ \KwTo $n - 1$}{
    \tcp{if it’s a last token, then skip it and encode the metacharacter}
    \uIf{$f[T[i]] > 0$} { 
        $w \leftarrow T[i]$\;
    }
    \Else{
        $w \leftarrow \text{lt}$\; 
        $l \leftarrow l - 1$\;
    }

    \While{$x \ge 2 \cdot f[w]$}{
        append least significant bit of $x$ to $\text{Code}$\;
        $x \leftarrow \lfloor x / 2 \rfloor$\;
    }

    $x \leftarrow l + I_w[x - f[w]]$\;

    $l \leftarrow l - 1$\;
    $f[w] \leftarrow f[w] - 1$\;
}

\Return $\text{Code}, W_0$\;

\caption{tANS with FAM modification}
\end{algorithm}

It is worth to say that here the table is partitioned onto multiple sub-tables corresponding to each symbol, while on practice one can join them using the accumulative frequencies.
The deciding algorithm is mirrored with a distinction, where the table is being built in the main loop based on the following properties. The state $x$ is known after the last step and equals $1$. At each step of the loop the inequality $x - l \leq l -1$ holds; in particular, if $x - l < l - 1$, then the difference $x - l$ equals the position of a decoded character; otherwise, the current encoded symbol is a metachareacter $lt$.
The algorithm has been implemented on C++. The compression process is divided into several steps:

\begin{itemize}
    \item reading the text from a file;
    \item text tokenization;
    \item dictionary building and sorting, token indexation, and counting;
    \item the algorithm itself with bits flush into an output stream;
    \item separate dictionary file creation and its compression.
\end{itemize}

\section{Experiments and results}

The proposed algorithm had been compared with two well-known ways to shuffle indices: rANS and almost uniform permutation (the classical tANS). In the first case, the table is built faster, while the second one produces a shorter code. In addition, we experimented with the shuffling, which is identical to the one in our method. All approaches allow to order the dictionary lexicographically, which can improve its compression. \\
For all algorithms, we used word-wise tokenization, disregarding letter cases and punctuation in front of the words. The dictionary and frequencies were compressed with the 7z compressor using the algorithm PPMd with the maximum compression level \cite{ppmd}. For the experiments, we used the following texts: dickens and webster from the Silesia corpus, book1 and book2 from the Calgary corpus, and alice29 and asyoulik from the Canterbury corpus. The results of the experiments are shown in Tables \ref{t1} and \ref{t2}. The sizes of all files are measured in bytes and execution time is in seconds. We should clarify that the depicted time includes only indexed text compression, without input preprocessing. The experiments were conducted on a device with a 12th Gen Intel(R) Core(TM) i5-12500H CPU. \\
One can see that the code size generated by our method is significantly lower than the other ones. Our algorithm execution time slightly varies from the ranged ANS, while the uniform permutation shows the greatest one, because its table construction involves a priority queue of a large alphabet size being updated at each step. Although the lexicographically ordered dictionary has been compressed better, it does not provide a superior benefit in terms of total size.

\begin{table}[h]
\caption{Dictionary and frequencies compressed file sizes}
\label{t1}
\centering
\begin{tabular}{|c|c|c|c|}
\toprule
Text &
Last appearance ordered dictionary &
Lexicographically ordered dictionary &
Frequencies \\
\midrule

dickens
& 321153 & 314282 & 42847 \\

webster
& 2708702 & 2483657 & 146873 \\

book1
& 76161 & 72152 & 7824 \\

book2
& 55129 & 53560 & 6354 \\

alice29
& 19160 & 18381 & 2302 \\

asyoulik
& 18554 & 18173 & 2067 \\

\bottomrule
\end{tabular}
\end{table}

\begin{table}[h]
\caption{Experiments results}
\label{t2}
\centering
\small
\resizebox{\columnwidth}{!}{\begin{tabular}{|l|ccc|ccc|ccc|ccc|}
\toprule
\multirow{2}{*}{Text} &
\multicolumn{3}{c|}{Forward Coding} &
\multicolumn{3}{c|}{Ranged permutation} &
\multicolumn{3}{c|}{Uniform permutation} &
\multicolumn{3}{c|}{Reversed permutation} \\
\cmidrule(lr){2-4} \cmidrule(lr){5-7} \cmidrule(lr){8-10} \cmidrule(lr){11-13}
& Code & Tot. & Time
& Code & Tot. & Time
& Code & Tot. & Time
& Code & Tot. & Time \\
\midrule

dickens
& 2888357 & 3209510 & 0.6209
& 3051018 & 3408147	& 0.3567
& 3035330 & 3392459 & 10.4548
& 3035826 & 3392955 & 0.6245 \\

webster
& 7438759 & 10147461 & 2.1179
& 8859441 &	11489971 & 1.3064
& 8780665 & 11411195 & 25.0070
& 8779496 & 11410026 & 2.0174 \\

book1
& 199140 & 275301 & 0.0633
& 232005 & 311981 & 0.0368
& 230688 & 310664 & 0.6149
& 230657 & 310633 & 0.0425 \\

book2
& 149691 & 204820 & 0.0356
& 171835 & 231749 & 0.0293
& 170058 & 229972 & 0.4709
& 170052 & 229966 & 0.0330\\

alice29
& 34798 & 53958 & 0.0114
& 41725 & 62408 & 0.0074
& 41491 & 62174 & 0.1226
& 41456 & 62139 & 0.0117 \\

asyoulik
& 30218 & 48772 & 0.0111
& 37361 & 57601	& 0.0099
& 36980 & 57220 & 0.1093
& 36967 & 57207 & 0.0132\\

\bottomrule
\end{tabular}}
\end{table}

\section{Conclusion}

It is shown that FAM allows to implement an adaptive version of compression based on ANS, which has better results in word based natural language text compression, than the original variant. Next research may include application of the proposed method for arbitrary data format compression.

%\subsection{Headings: second level}

%\subsubsection{Headings: third level}

%\paragraph{Paragraph}

%The documentation for \verb+natbib+ may be found at
%\begin{center}
%  \url{http://mirrors.ctan.org/macros/latex/contrib/natbib/natnotes.pdf}
%\end{center}
%Of note is the command \verb+\citet+, which produces citations
%appropriate for use in inline text.  For example,
%\begin{verbatim}
%   \citet{hasselmo} investigated\dots
%\end{verbatim}
%produces
%\begin{quote}
%  Hasselmo, et al.\ (1995) investigated\dots
%\end{quote}

%\begin{center}
%  \url{https://www.ctan.org/pkg/booktabs}
%\end{center}

%\subsection{Figures}
%\lipsum[10] 
%See Figure \ref{fig:fig1}. Here is how you add footnotes. \footnote{Sample %of the first footnote.}
%\lipsum[11] 

%\begin{figure}
%  \centering
%  \fbox{\rule[-.5cm]{4cm}{4cm} \rule[-.5cm]{4cm}{0cm}}
%  \caption{Sample figure caption.}
%  \label{fig:fig1}
%\end{figure}

%\subsection{Lists}

%Bibliography


\begin{thebibliography}{9}

\bibitem{rissanen1976}
J. J. Rissanen,
``Generalized Kraft inequality and arithmetic coding,''
\textit{IBM Journal of Research and Development},
vol. 20, no. 3, pp. 198--203, May 1976.

\bibitem{duda2009}
J. Duda,
``Asymmetric Numerical Systems,''
arXiv preprint arXiv:0902.0271, 2009.
[Online]. Available: \url{https://arxiv.org/pdf/0902.0271}

\bibitem{duda2014}
J. Duda,
``Asymmetric Numeral Systems: Entropy Coding Combining Speed of Huffman Coding with Compression Rate of Arithmetic Coding,''
arXiv preprint arXiv:1311.2540, 2014.
[Online]. Available: \url{https://arxiv.org/pdf/1311.2540}

\bibitem{zavadskyi2024}
I. O. Zavadskyi, S. T. Klein, and D. Shapira,
``Word-based Forward Coding,''
in \textit{Data Compression Conference (DCC)}, 2024, pp. 352--361.

\bibitem{zavadskyi2026}
I. O. Zavadskyi and D. Shapira,
``Forward Modeling in Adaptive Compression: Bounds and Experimental Evaluation,''
in \textit{Data Compression Conference (DCC)}, 2026, pp. 223--232.

\bibitem{ppmd}
``PPMd Compression,''
Mintlify.wiki, 2026.
[Online]. Available: \url{https://mintlify.wiki/ip7z/7zip/compression/ppmd}
[Accessed: Apr. 18, 2026].

\end{thebibliography}
\end{document}